\documentclass[twoside,11pt]{amsart}
\usepackage{amsmath}  
\newcommand{\new}{\newcommand} 
\new{\bg}{\begin}

\theoremstyle{remark}
\newtheorem{remark}{Remark}

\new{\iii}{\begin{enumerate}} 
\new{\fff}{\end{enumerate}}
\new{\mfi}{\begin{eqnarray*}} 
\new{\mff}{\end{eqnarray*}} 
\new{\mfni}{\begin{eqnarray}}
\new{\mfnf}{\end{eqnarray}}
\new{\be}[2]{\begin{equation}\label{#1}{#2}\end{equation}}
\new{\eqn}[1]{~(\ref{#1})}
\new{\nor}[1]{\left\|{#1}\right\|}
\new{\nori}[1]{\left\|{#1}\right\|_{\sup}}
\new{\scal}[2]{\left\langle{#1},{#2}\right\rangle}
\new{\room}{\ \ \ \ }
\new{\set}[1]{\{{#1}\}} 
\new{\hh}{{\mathcal H}}
\new{\runo}{{\mathbb R}}
\new{\cuno}{{\mathbb C}}
\new{\m}{^{-1}}
\new{\dom}{\mathrm{dom}\,}
\new{\lie}{\mathfrak{H}}
\new{\lieg}{\mathfrak{G}}
\new{\intg}[2]{\int_{G}{#1}\,d#2}
\new{\inth}[2]{\int_{H}{#1}\,d#2}
\new{\intk}[2]{\int_{K}{#1}\,d#2}
\new{\intl}[2]{\int_{\lie}{#1}\,d#2}
\new{\hs}{{\mathcal L}^2(\hh)}
\new{\ld}{L^2(H,dh)}
\new{\Tr}[1]{\mathrm{Tr}\,(#1)}
\new{\ddd}{S^m}
\new{\pippo}{{\mathcal L}({\cuno^{2j+1}})}
\new{\q}{a}
\new{\p}{b}
\new{\ac}{\hat a}
\new{\cc}{\hat a^\dag}

\begin{document}

\title{Group Theoretical Quantum Tomography}
\author{G. Cassinelli}
\address{Gianni Cassinelli, Dipartimento di Fisica,
Universit\`a di Genova, I.N.F.N., Sezione di Genova, Via Dodecaneso~33,
16146 Genova, Italy}
\email{cassinelli@ge.infn.it}
\author{G.M. D'Ariano}
\address{Giacomo Mauro D'Ariano, Dipartimento di Fisica 'Alessandro
Volta', Universit\`a di Pavia, I.N.F.M., Unit\`a di Pavia, Via
Bassi~6, I-27100 Pavia, Italy}
\email{dariano@pv.infn.it}
\author{E. De Vito}
\address{Ernesto De Vito, Dipartimento di Matematica, Universit\`a di
Modena, Via Campi 213/B, 41100 Modena, Italy and I.N.F.N., Sezione di Genova,
Via Dodecaneso~33, 16146 Genova, Italy}
\email{devito@unimo.it}
\author{A. Levrero} 
\address{Alberto Levrero, Dipartimento di Fisica, Universit\`a di
Genova, I.N.F.N.,  Sezione di Genova, Via Dodecaneso~33, 16146 Genova,
Italy}
\email{levrero@ge.infn.it}
\date{\today}

\begin{abstract}
The paper is devoted to the mathematical foundation of the quantum
tomography using the theory of square-integrable representations of unimodular Lie groups.
\end{abstract}

\maketitle

\section{introduction}
In Quantum Mechanics a physical system is associated with a Hilbert
space $\hh$: the states are described by positive trace-class 
trace-one operators $T$ on $\hh$, the physical quantities by self-adjoint
operators $A$ on $\hh$ and the physical content of the theory is given
by the expectation values $\Tr{AT}$. 
The state $T$ is completely determined by $\Tr{Q_nT}$ for $Q_n$ running 
on a suitable set $\set{Q_n}$ of observables and, for arbitrary operator $A$, $\Tr{AT}$ can be
computed in terms of $\Tr{Q_nT}$.
In order to implement this scheme one has to  estimate  $\Tr{Q_nT}$
experimentally, facing the problems arising from statistical errors and
instrumental noise. Moreover, the number of experimental data is clearly finite,
while $A$ and $T$ are operators on an infinite dimensional Hilbert space and
the set $\set{Q_n}$ is infinite.

The problem of determining the state of a quantum system entered the realm of
experiments in the last decade, in the domain of quantum optics. Many authors,
see for example \cite{chi,smithey,mauro1,bre}, proposed and used various
techniques to reconstruct the density operator of a single mode of the e.m.
field from the probability distributions of its quadratures. These methods
were originally based on the use of the Radon transform, as in medical
tomographic imaging.  Due to this analogy the name {\em quantum tomography}
is currently used to refer to these techniques.
Their common feature, for a review see \cite{R4}, is the use of a set of observables
$\set{Q_n\ : \ n\in X}$, called {\em quorum}, parametrised by a space $X$
endowed with a probability measure $\mu$. The fundamental property of the
quorum is that any observable $A$ can be expressed as {\em integral transform}
on the space $X$
$$A=\int_X \mathcal{E}[A](n)\ d\mu(n)$$
in such a way that, for all
$n\in X$, the operator $\mathcal{E}[A](n)$ is a function of $Q_n$ in
the sense of the functional calculus. Then, if $T$ is the
state, one has that
\be{zero}{\Tr{AT} = \int_{X\times\runo} \sigma(A)(n,\lambda)\ 
\omega(n,\lambda)\ d\mu(n)d\lambda,}
where  $\lambda\mapsto\omega(n,\lambda)$ is the probability density of $Q_n$ in the
state $T$, {\em i.e.}
$$\Tr{TQ_n} = \int_{\runo} \lambda\ \omega(n,\lambda)\ d\lambda,$$
and $\lambda\mapsto\sigma(A)(n,\lambda)$ is the function defined by
$\mathcal{E}[A](n)$ using the functional calculus, {\em i.e.} 
$$\Tr{T\mathcal{E}[A](n)} = \int_{\runo} \sigma(A)(n,\lambda)
\omega(n,\lambda)\ d\lambda$$
(in the above formulas we assumed for simplicity
that each $Q_n$ has a continuous spectrum). Selecting randomly $Q_n$ in the
quorum according to the probability measure $\mu$ and measuring it, the
outcome probability of obtaining the value $\lambda$ is given by
$\omega(n,\lambda)d\mu(n)d\lambda$. Then, by means of Eq.\eqn{zero}, the
expectation value $\Tr{AT}$ can be reconstructed, by averaging the function
$\sigma(A)$ over $X\times\runo$ endowed with the probability measure $\omega
d\mu d\lambda$.  We notice that the function $\sigma(A)$, called the {\em
  estimator} of $A$, does not depend on $T$, and that the same set of data can
be used to estimate all the expectation values $\Tr{AT}$.

In \cite{R7} and \cite{paini-tesi} a general method has been proposed to
realize a quorum and define estimators in terms of suitable unitary
representations of Lie groups (for a self contained synthetic exposition see
\cite{mauro} and \cite{paini-rete}). The present paper is devoted to the
mathematical foundation of this method using the theory of square-integrable
representations of unimodular Lie groups. In Section~2 we present the
mathematical theory and in Section~3 we apply it to two examples: the homodyne
tomography related to the Weyl-Heisenberg group and the angular momentum
tomography associated with the rotation group.

\section{Group-dynamical quorum}
In this section we define a quorum associated with a square-integrable
representation of a Lie group.

Let $G$ be a unimodular connected Lie group $G$ and $K$ a central closed
subgroup. The quotient space $H=G/K$
is a unimodular connected Lie group. We denote by $\lie $ its Lie
algebra, by $m+1$ the (real) dimension of $\lie$ as a vector space, by
$dv$ the Lebesgue measure on $\lie$ and by $dh$ the Haar measure of $H$,
uniquely defined up to a positive constant, which will be fixed in the
following. 

Denoted by $\exp$ the exponential map from $\lie$ to $H$, we assume that there
is an open subset $V$ of $\lie$ such that $\exp(V)$ is open in $H$, its
complement has zero measure with respect to $dh$ and $\exp$ is a
diffeomorphism from $V$ onto $\exp(V)$.  This hypothesis implies that, given
$f\in L^1(H,dh)$, \be{lie}{\inth{f(h)}{h} = D
  \intl{f(\exp(v))|\mathrm{det\,}(d(\exp)_v)|\chi_V(v) }{v},} where
$d(\exp)_v$ is the differential of the exponential map at $v\in\lie$, {\em
  i.e.}
$$d(\exp)_v(w) = \left(\frac{d}{dt}\exp(-v)\exp(v+tw)\right)_{t=0}\room w\in\lie,$$
$\mathrm{det\,}(\cdot)$ is the determinant and $D$ is a positive
constant, see for example Th.~1.14, Ch.~I of \cite{helgason2}. 
We normalize the Haar measure $dh$ of $H$ in such a way that $D=1$.
\bg{remark}
The density $\mathrm{det\,}(d(\exp)_v)$ can be easily computed observing
that, if $\lambda_1,...,\lambda_{m+1}$ are the (possibly repeated) eigenvalues of $d(\exp)_v$,
viewed as linear operator on $\lie$, then 
$$\mathrm{det\,}(d(\exp)_v)= \frac{1 - e^{-\lambda_1}}{\lambda_1} \ldots 
\frac{1 - e^{-\lambda_{m+1}}}{\lambda_{m+1}},$$  
with $\frac{1 - e^{-0}}{0}=1,$ see for example Th.~1.7, Ch.~I of \cite{helgason}.
\end{remark}
Let $U$ be an irreducible continuous unitary representation of $G$. We 
denote by $\hh$ the (complex separable) Hilbert space where the representation acts and by 
$\scal{\cdot}{\cdot}$ the scalar product, linear in the second
argument. 

We assume that the representation $U$ is square-integrable modulo
$K$, {\em i.e.} there is a non-zero vector $v\in\hh$ such that
\be{square}{\inth{|\scal{U_{c(h)}v}{v}|^2}{h}<\infty,}
where $c$ is a section from $H$ to $G$, {\em i.e.} a  measurable map $c:H\to G$ such that
\mfi 
c(e_H) & = & e_G \\
\pi(c(h)) & =  & h\room h\in H, 
\mff
with $\pi$ being the canonical projection from $G$ to $H$.
Notice that the value of the integral in Eq.\eqn{square} is
independent of the choice of the section and that Eq.\eqn{square} implies that 
the function $h\mapsto \scal{U_{c(h)}u}{w}$ is square integrable for all
$u,w\in\hh$, \cite{borel}.
  
\bg{remark} 
In many examples $K$ is trivial, {\em i.e.} $K={e_G}$, so that $H=G$
and Eq.\eqn{square} reduces to the usual notion of square-integrability.
Nevertheless, there are cases, as the Weyl-Heisenberg
group, that require the full theory. Moreover,  in this framework one can
easily consider projective representations. Indeed, let $\widehat U$ be a
projective representation of a Lie group $\widehat H$ with multiplier $m$. Define $G$ as the
central extension of the torus $K$ by $\widehat H$ associated with $m$.
Then $K$ is a central closed subgroup of $G$, $H$ is canonically
isomorphic with $\widehat H$ and there is a unitary representation $U$ of $G$
such that
$$\widehat U_{\pi(g)} = U_g\room g\in G.$$
Clearly, the fact that $U$ is square-integrable modulo $K$ is
equivalent to the fact that $\widehat U$ is a square-integrable projective
representation of $\widehat H$.  
\end{remark}
Being $U$ square-integrable modulo $K$, one can prove \cite{borel}
that there is a constant $d>0$, called {\em the formal degree} of $U$, such that,  for all
$u_1,u_2,v_1,v_2\in\hh$, 
\be{ortho}{\inth{\overline{\scal{U_{c(h)}v_1}{u_1}}\scal{U_{c(h)}v_2}{u_2}}{h}=
{\frac{1}{d}}\scal{u_1}{u_2}\scal{v_2}{v_1}.} 

Using the above relation we can represent the Hilbert-Schmidt operators as
square integrable functions on $H$.
Indeed, let $\hs$ be the Hilbert space of the Hilbert-Schmidt operators
with the scalar product
$$(A,B)\mapsto\Tr{A^*B},$$
where $\Tr{\cdot}$ denotes the trace and $A^*$ is the adjoint operator of $A$.
If $u,v\in\hh$, let $u\otimes v^*$ be the operator in $\hs$
$$(u\otimes v^*)(w) = \scal{v}{w}u\room w\in\hh.$$
Given a section $c$, we define $\Sigma(u\otimes v^*)$ as
the function from $H$ to $\cuno$ given by
$$\Sigma(u\otimes v^*)(h) = \scal{U_{c(h)}v}{u}\room h\in H.$$
From Eq.\eqn{ortho}, it follows that $\Sigma(u\otimes v^*)$ is square-integrable with
respect to $dh$ and 
$$\nor{\Sigma(u\otimes v^*)}_{\ld}^2=\frac{1}{d}\nor{u}^2\nor{v}^2=
\frac{1}{d}\nor{u\otimes v^*}^2_{\hs}.$$
Taking into account that the set 
$\set{u\otimes v^* \ : \ u,v\in\hh}$ is total in $\hs$, it follows
that $\Sigma$ is defined uniquely by continuity on $\hs$ and, if $A,B\in\hs$, 
\be{iso}{\Tr{A^*B} = d \scal{\Sigma(A)}{\Sigma(B)}.}  
Moreover, if $A$ is of trace-class, then for almost all $h\in H$
\be{trace}{\Sigma(A)(h)= \Tr{U_{c(h)\m }A}.} 
Indeed, let 
$$A=\sum_i \lambda_i e_i\otimes f_i^*$$
be the canonical decomposition of $A$, where $(e_i)$ and $(f_i)$ are 
orthonormal sequences in $\hh$, $(\lambda_i)$ is an
$\ell_1$-sequence and the series converges in trace-norm and, hence,
in the Hilbert-Schmidt norm. Since $\Sigma$ is continuous, then
$$\Sigma(A)  = \sum_i \lambda_i \Sigma(e_i\otimes f_i^* ),$$
where the series converges in $\ld$. On the other hand, fixed $h\in H$,
since $A$ is of trace class, so is $U_{c(h)\m}A$, hence
\mfi
\Tr{U_{c(h)\m}A} & = & \sum_{i} \scal{f_i}{U_{c(h)\m}Ae_i} \\
& = & \sum_{i}\lambda_i \scal{U_{c(h)}f_i}{e_i} \\
& = & \sum_i \lambda_i \Sigma(e_i\otimes f_i^* )(h),
\mff
where the series converges pointwise. The claim is now clear.

We are now ready to define a quorum associated with the
square-integrable (modulo $K$) representation $U$ of $G$.

Let $T$ be  a state of $\hh$, {\em i.e.} a positive trace-class operator 
of trace one, and $A$ a
Hilbert-Schmidt operator on $\hh$. Taking into account
Eq.\eqn{iso} and Eq.\eqn{trace},
\mfi
\Tr{TA}  & = & d \scal{\Sigma(T)}{\Sigma(A)}_{\ld} \\
& = & d \inth{\overline{\Tr{U_{c(h)\m}T}}\Sigma(A)(h) }{h},  
\mff
so that 
$$\Tr{AT} = d \inth{\Sigma(A)(h) \Tr{TU_{c(h)}}}{h}.$$
By means of Eq.\eqn{lie}, the above equation becomes
$$ \Tr{AT} =d
\intl{\Sigma(A)(\exp{v})\Tr{TU_{c(\exp{v})}}\chi_V(v)|\mathrm{det\,}(d(\exp)_v)|}{v}.$$
Let $S^m$ be the sphere in $\lie$. Then, for all $n\in S^m$, the map
$$t\mapsto U_{c(\exp{(t n)})}$$
is a projective representation of $\runo$. Since all the multipliers of
$\runo$ are equivalent to an exact one,  there is a selfadjoint
unbounded operator $Q_n $ and a measurable complex function $\alpha_n$
with modulo 1 such that, for all $t\in\runo$,
\be{quorum}{ U_{c(\exp{(t n)})}= \alpha_n(t) e^{it Q_n }.}
Using polar coordinates in the above equation, one has that 
\mfni
\Tr{AT} & = & d C_m \int_{\ddd}d\Omega(n)\int_0^{\infty}  dt\ 
t^m \Sigma(A)(\exp{(t n)})\alpha_n(t)\label{8bis} \\
& & \room\room\Tr{Te^{it Q_n }}\chi_V(tn)|\mathrm{det\,} (d(\exp)_{t n})|,\nonumber
\mfnf
where $d\Omega$ is  the normalized measure on the sphere $S^m$, $C_m$
is the volume of $S^m$ and $dt$ is the Lebesgue measure on the real line.
The set of self-adjoint operators $\set{Q_n\ : n\in\ddd}$, labelled by the
probability space $(\ddd,d\Omega)$, is called the {\em quorum}
defined by the representation $U$. We notice that Eq.\eqn{quorum}
defines $Q_n$ uniquely up to an additive constant, see, also,
Remark~\ref{furba} below.

Since $Q_n $ is selfadjoint, by means of the spectral theorem, 
there is a projection valued measure $E\mapsto P_n(E)$ defined on
$\runo$ such that 
$$\Tr{TQ_n}  = \int_{\runo} \lambda \ d\Tr{TP_n(\lambda)},$$
where $d\Tr{TP_n(\lambda)}$ denotes the positive bounded measure on $\runo$
$$E\mapsto\Tr{TP_n(E)}.$$

Using this equation, one obtains that 
\mfni 
\Tr{AT} & = & d C_m
\int_{\ddd}d\Omega(n)\int_0^{\infty}dt \int_{\runo} 
d\Tr{TP_n(\lambda)}\label{su}\\
& & \ \ \ \ e^{i\lambda t} \Sigma(A)(\exp{(t n)})\alpha_n(t)
\chi_V(tn)|\mathrm{det\,} (d(\exp)_{t n})| t^m.\nonumber  
\mfnf
In order to obtain a reconstruction formula for $\Tr{AT}$, we would like
to interchange the integrals in $dt$ and in $d\Tr{TP_n(\lambda)}$.

We consider first the case  when $\Sigma(A)$, which is only
square-integrable, is in fact integrable with respect to $dh$, {\em i.e.}
\be{condi}{\inth{|\Sigma(A)(h)|}{h}<\infty.} 
By means of Fubini theorem, this condition implies that, for almost all $n\in\ddd$, the map
$t\mapsto\Sigma(A)(\exp{(tn)})$ is integrable with respect to
the measure
\be{tn}{dt_n=\chi_V(tn)|\mathrm{det\,} (d(\exp)_{t n})| t^m dt.}
Then the map from $\ddd\times\runo$ to $\cuno$
\be{defi1}{\sigma(A)(n,\lambda) ={dC_m} \int_0^{\infty} e^{i\lambda t}
\Sigma(A)(\exp{(tn)})\alpha_n(t) \chi_V(tn)|\mathrm{det\,} (d(\exp)_{t n})|  t^m\ dt,}
is well-defined and it is called the {\em  estimator} of the
observable $A$. We notice that the estimator  does not depend on $T$ and, given the
representation $U$, can be computed analytically.

Since the measure $d\Tr{TP_n(\lambda)}$ is bounded, by means of Fubini theorem,
one can interchange  the integrals in Eq.\eqn{su} obtaining
\be{final}{\Tr{AT} = \int_{\ddd}d\Omega(n)\int_0^{\infty}
d\Tr{TP_n(\lambda)}\ \sigma(A)(n,\lambda).}
The above integral transform is the core of the {\em quantum
tomography} and is a concrete realization of the scheme proposed in
the introduction, compare with Eq.\eqn{zero}. Indeed, $d\Omega(n)d\Tr{TP_n(\lambda)}$ is the
probability  to obtain the value $\lambda$ when the observable $Q_n$,
chosen randomly in the quorum according to $d\Omega$, is
measured.
Moreover,  by means of Eq.\eqn{final},
the expectation value $\Tr{AT}$ can be reconstructed as average of
the estimator $\sigma(A)$ over many random measures of the observables $Q_n$
in the quorum.

\bg{remark}\label{furba}
There is a  choice for the section that 
simplifies the expression of the estimator. Indeed, denoted by $\lieg$ the Lie algebra of $G$,
since the differential $d\pi$ of $\pi$ is a surjective linear map from
$\lieg$ onto $\lie$, there is an injective linear map $j$ from $\lie$ to
$\lieg$ such that $d\pi(j(v))=v$ for all $v\in\lie$. Since $\exp$ is a
diffeomorphism from $V$ onto $\exp(V)$, it is well defined a smooth  
map $\hat c$ from $\exp(V)$ to
$G$ such that
$$ \hat c(\exp(v)) = \exp(j(v))\room v\in\lie.$$
Clearly $\hat c$ is a section and the relation
$U_{ \hat c(\exp(tn)) }= U_{\exp{(tj(n))}}$ shows that one can always choose 
$\alpha_n(t)= 1$ in Eq.\eqn{quorum}. Hence $U_{ \hat c(\exp(tn)) } = e^{it Q_n}$.

One can easily prove that, if one changes $j\mapsto j+l$ in such a way
that $d\pi(j(v)+l(v))=v$, then the quorum transforms according to
$Q_n\mapsto Q_n+ q_n I$. 
However, in most of the cases, there is a {\em natural} choice for the map $j$, so
that the quorum $Q_n$ is, in fact, defined uniquely by the representation $U$.
\end{remark}
\bg{remark}
Once the quorum $\set{Q_n}$ is fixed, Eq.\eqn{defi1} is 
independent of the choice of the section $c$. Indeed if $c'$ is
another section, then, for all $h\in H$, $c'(h)=k(h)c(h)$ and $k(h)\in
K$. Since $K$ is central in $G$ and $U$ is irreducible, then
$U_{k(h)}=\beta(h)I$, where $\beta(h)$ is a complex number of modulo
one. Hence, with obvious notations, for almost all $h\in H$ and for all $t\in\runo$,
\mfi
\Sigma'(A)(h) & = & \overline{\beta(h)} \Sigma(A)(h) \\
{\alpha'}_n(t) & = & \beta(h) \alpha_n(t),
\mff
so that $\sigma(A)$ is invariant with respect to the change
$c\mapsto c'$. 
\end{remark}
\bg{remark}
If $A$ is of trace class and satisfies Eq.\eqn{condi}, using Eq.\eqn{trace} one obtains a more explicit
formula for the estimator of $A$
$$\sigma(A)(n,\lambda) = dC_m \int_0^{\infty} e^{i\lambda t}
\Tr{Ae^{-itQ_n}} \chi_V(tn)|\mathrm{det\,} (d(\exp)_{t n})|  t^m\ dt.$$
Moreover, in most examples the set $V$ is sufficiently nice so that
the map $n\mapsto\chi_V(tn)$ is continuous for almost all $t\in\runo$. 
In this case, if one
chooses the section $\hat c$ as in Remark~\ref{furba}, taking into account that
the function $g\mapsto \Tr{TU_g}$ is continuous (since the ultra-weak
operator topology is equivalent to the weak operator topology on the
unit ball of $\mathcal{L}(\hh)$), it follows that the estimator
$\sigma(A)$ is continuous on $\ddd\times\runo$. This 
property is important in order to approximate the integral of Eq.\eqn{final} 
by a finite sum.
\end{remark} 
\bg{remark}
We notice that this procedure is {\em unbiased} since the observables
$Q_n$ are chosen randomly and the integral given by Eq.\eqn{final}
can be approximated by a finite sum since
$d\Omega(n)d\Tr{TP_n(\lambda)}$ is a probability measure. 
This means that this approach is not affected by the systematic errors
that were present in the first tomographic scheme \cite{chi}, \cite{smithey}
due to the cutoff needed in the inversion of the Radon transform, see \cite{mauro1}.   
\end{remark} 

\bg{remark}
If $H$ is compact then  $dh$ is finite and any irreducible
representation  is square-integrable. Since the Hilbert space
$\hh$ where the representation acts is finite dimensional, $\hs$
coincides with the space of the all operators.
Moreover, since $\ld \subset L^1(H,dh)$, Eq.\eqn{condi}
holds for every operator. 
\end{remark}

\begin{remark} 
  If $U$ is an integrable representation (modulo $K$), there
  exists a dense set $S$ in $\hh$ such that, if $u,v\in S$, then
  $\Sigma(u\otimes v^*)$ satisfies Eq.\eqn{condi}.
\end{remark}

If condition \eqn{condi} does not hold, it may  happen that, for a non-negligible set of
$n\in\ddd$,  the map $t\mapsto   \Sigma(A)(\exp{(t n)}) $ is not
integrable with respect to the measure $dt_n$ defined by Eq.\eqn{tn} (it is
only square-integrable), so that the estimator $\sigma(A)$ given by
Eq.\eqn{defi1} is not well defined.

In these cases, in order to define the estimator one has to use a suitable
regularization procedure. For example, fixed $L>O$ let, for all $n\in\ddd$
and $\lambda\in\runo$,
\be{defi2}{\sigma_L(A)(n,\lambda) =dC_m \int_{0}^{L} e^{i\lambda t}
\Sigma(A)(\exp{(tn)})\alpha_n(t) \chi_V(tn)|\mathrm{det\,} (d(\exp)_{t n})|  t^m\ dt.}
It may be the case that there exists a function $\sigma(A)$ such
that 
\mfi
&& \lim_{L\to\infty} \int_{\ddd}d\Omega(n)\int_{\runo} d\Tr{TP_n(\lambda)}
\ \sigma_L(A)(n,\lambda)\room\room  = \\
&& \room\room\room \int_{\ddd}d\Omega(n)\int_{\runo} d\Tr{TP_n(\lambda)}
\ \sigma(A)(n,\lambda).
\mff
Then, as an easy consequence of dominated convergence theorem, one has that
$$
{\Tr{AT} =\int_{\ddd}d\Omega(n)\int_{\runo} d\Tr{TP_n(\lambda)}
\ \sigma(A)(n,\lambda).}
$$
Analogous regularization procedures could be used to extend $\Sigma(A)$ to
non-Hilbert-Schmidt operators. Although this problem is physically relevant
(many observables of interest are unbounded) it is out of the scope of the
present paper.

\section{examples}

\subsection{The Weyl-Heisenberg group }
Let $G$ be the Weyl-Heisenberg group, {\em i.e.} $G=\runo^3$ with the
composition law
$$(\eta_1,\q_1,\p_1)(\eta_2,\q_2,\p_2) =
(\eta_1+\eta_2 +\frac{\p_1\q_2-\q_1\p_2}{2},\q_1+\q_2,\p_1+\p_2).$$
It is known that $G$ is a connected simply-connected nilpotent
unimodular Lie group.

The set $K=\set{(\eta,0,0)\ : \ \eta\in\runo}$ is clearly a central
closed subgroup of $G$ and the quotient group $H=G/K$ can be identified with the vector
group $\runo^2$. One has the following facts.
\iii
\item The canonical projection $\pi$ is given by $\pi(\eta,\q,\p)=(\q,\p)$.
\item A smooth section $c$ is given by $c(\q,\p)=(0,\q,\p)$.
\item A Haar measure on $H$ is the Lebesgue measure $d\q \,d\p$ of
  $\runo^2$.
\item The Lie algebra $\lie$ of $H$ can be identified with $\runo^2$ so
that the exponential map is the identity and, for all $v\in\lie$,
$\mathrm{det\,} (d(\exp)_{v})=1$.
\item The constant $D$ in Eq.\eqn{lie} is equal to $1$.
\fff
It follows that the choice  $V=\lie$ satisfies the assumptions of the
previous section.

Let $U$ be the representation of $G$ acting in $\hh=L^2(\runo,dx)$ as
$$(U_{(\eta,\q,\p)}u)(x) = e^{i(\eta+\frac{\q\p}{2})} e^{ix\q} u(x+\p)$$
where $x\in\runo$, $u\in L^2(\runo,dx)$ and $(\eta,\q,\p)\in G$. It is
known that $U$ is a unitary continuous irreducible representation of
$G$, called the {\em Schr\"odinger representation}.

We prove that $U$ is square-integrable modulo $K$.
Given $u\in\hh$, the map
$$ (x,\p)\mapsto \overline{u(x+\p)}u(x)$$
is measurable and
$$\int_{\runo} dx\int_{\runo} d\p  |\overline{u(x+\p)}u(x)|^2 =\nor{u}^4<\infty.$$
By Fubini theorem, for almost all $\p\in\runo$, the map
$$x\mapsto \overline{u(x+\p)}u(x)$$
is square-integrable and, since both $x\mapsto u(x+\p)$ and $x\mapsto
u(x)$ are square-integrable, it is integrable. Then, the map
$$\q\mapsto\int_{\runo}dx e^{-i\q x}\overline{u(x+\p)}u(x)$$
is well defined and square-integrable with respect to $d\q$ for almost all $\p\in\runo$.
Moreover, the map
$$(\q,\p)\mapsto\int_{\runo}dx e^{-i\q x}\overline{u(x+\p)}u(x)$$
is measurable and, by means of the isometry of the Fourier transform,
\mfi
\int_{\runo} d\p\int_{\runo} d\q\left|\int_{\runo}dx
e^{-i\q x}\overline{u(x+\p)}u(x)\right|^2 & = & 2\pi
\int_{\runo} d\p \int_{\runo}dx |\overline{u(x+\p)}u(x)|^2 \\
& = & 2\pi \nor{u}^4.
\mff
Since
$$\scal{U_{c(\q,\p)}u}{u} = \int_{\runo}dx e^{-i\q x}\overline{u(x+\p)}u(x),$$
by Fubini theorem one has that $U$ is square-integrable and the formal degree is $d=\frac{1}{2\pi}$.

Since $U$ is square-integrable modulo $K$, according to Section~2, it defines a quorum. In order
to explicit it, we observe that,  with the notation of the previous
section,
$$\ddd=\set{n_{\Phi}:=(\cos(\Phi), \sin(\Phi) )\ : \ \Phi\in[0,2\pi]},$$
$m=1$, $C_1=2\pi$ and $d\Omega=\frac{d\Phi}{2\pi}$. Moreover,
since $t\mapsto U_{c(t n_{\Phi})}$ is a one parameter subgroup, then
$$U_{c(t n_{\Phi})}= e^{it Y_{\Phi}}$$
where $Y_{\Phi}$ is a selfadjoint operator (in this example
$\alpha_{n_{\Phi}}(t)=1$). If $u$ is a Schwartz function, one has that
$$Y_{\Phi} u= \cos(\Phi) Qu + \sin(\Phi)Pu,$$ where $Q$ is the
multiplicative operator by $x$, {\em i.e.} the position operator, and
$P$ is $-i$ times the weak derivative operator, {\em i.e.} the momentum
operator. Hence the quorum defined by $U$ is given by the set of
self-adjoint operators
$$\set{Y_{\Phi} \ : \ \Phi\in [0,2\pi]}$$
labelled by the space $[0,2\pi]$ with the uniform measure $\frac{d\Phi}{2\pi}$.

The above quorum has the following property. For all $\Phi\in [0,2\pi]$, there is a
unitary operator $W_{\Phi}$ such that
\be{intre}{Y_{\Phi}=W_{\Phi}QW_{\Phi}\m.}
To prove it, given $\Phi\in [0,2\pi]$, let $f_\Phi$ from $G$ to $G$
$$f_\Phi(\eta,\q,\p) = \left(\eta, \cos(\Phi)\q-\sin(\Phi)\p, \sin(\Phi) \q
+\cos(\Phi) \p\right).$$
One can easily check that $f_\Phi$ is a continuous group automorphism of $G$, so
that $g\mapsto U^{f_{\Phi}}$ is a unitary irreducible continuous
representation of $G$ and the restriction to $K$ is the character
$\eta\mapsto e^{i\eta}$. From the unicity of the Schr\"odinger representation, it
follows that there exists a unitary operator $W_{\Phi}$ such that
$$U^{f_{\Phi}} = W_{\Phi} U W_{\Phi}\m.$$
Then
$$U_{c(tn_{\Phi})} = U^{f_{\Phi}}_{(0,t,0)} = W_{\Phi} U_{(0,t,0)}
W_{\Phi}\m, $$
and Eq.\eqn{intre} follows by Stone theorem.

Let now $T$ be a state of $\hh$. Recalling that the spectral
measure $P_{Q}$ of $Q$ is the one given by the multiplicative operators by characteristic
functions, then, by means of Eq.\eqn{intre}, 
for each $\Phi\in [0,2\pi]$ there is a $L^1(\runo,d\lambda)$-function $\lambda\mapsto \omega(\Phi,\lambda)$
such that
$$\Tr{TP_{\Phi}(E)} = \Tr{W_\Phi\m T W_\Phi P_{Q}(E)} = \int_E \omega(\Phi,\lambda) d\lambda,$$
where $E\mapsto P_{\Phi}(E)$ is the spectral measure associated with $Y_{\Phi}$.
The map  $\omega$ can  always be chosen  measurable 
as a function on $[0,2\pi]\times\runo$ and, in doing so, it is a probability density on
$[0,2\pi]\times\runo$ with respect to the measure $\frac{d\Phi}{2\pi}d\lambda$.

Finally, fix a Hilbert-Schmidt operator $A$ in $\hh$  such that $\Sigma(A)$ is integrable
with respect to $d\q\, d\p$. According to Eq.\eqn{defi1}, the estimator of
$A$ is, if $\Phi\in
[0,2\pi]$ and $\lambda\in\runo$,
$$\sigma(A)(\Phi,\lambda) = \int_0^{\infty}  t
\Sigma(A)(t\cos(\Phi),t\sin(\Phi))  e^{i\lambda t}\ dt,$$
and the reconstruction formula Eq.\eqn{final} is explicitly given by
$$ \Tr{AT}=\int_0^{2\pi}\int_\runo
\sigma(A)(\Phi,\lambda) \omega(\Phi,\lambda)\ \frac{d\Phi}{2\pi} d\lambda.$$ 

The representation $U$ is actually integrable and, if $(u_n)$  is the
basis of eigenvectors of the number operator, 
then $\Sigma(u_n\otimes u_{n+l}^*)\in L^1(H,dh)$ and one has the explicit formula 
\be{hyper}{
\sigma(u_n\otimes u_{n+l}^*)(\Phi,\lambda) = 
\frac{(-i)^{l}}{2^{\frac{l}{2}}}
\sqrt{\frac{n!}{(n+l)!}} e^{il\Phi}\int_0^{\infty} t^{l+1} L^{l}_{n}(\frac{t^2}{2})
e^{-\frac{t^2}{4}+i\lambda t} \ dt,} 
where $L^k_m$ are the associated Laguerre polynomials. 
The statistical reliability of Eq.\eqn{hyper} has been verified in
\cite{mauro1}.

This example is physically realized  by the
homodyne tomography, \cite{R4}. The quantum system is the harmonic
oscillator representing a single mode of the e.m. field with
annihilation and creation operators $\ac$ and $\cc$. In terms of such
operators, one has the following {\em translation table} 
\mfi
Q & = & \frac{\ac + \cc}{\sqrt{2}} \\
P & = & \frac{\ac - \cc}{\sqrt{2}i} \\
U_{(\eta,\q,\p)} & = & e^{i\eta} e^{\left(\alpha \cc -\overline{\alpha}\ac\right)} \\
Y_\Phi & = & \sqrt{2} \frac{\cc e^{i\Phi}+\ac e^{-i\Phi}}{2} =:\sqrt{2} X_\Phi
\mff
where $\alpha= \frac{-\p+i\q}{\sqrt{2}}\in\cuno$, $e^{\left(\alpha \cc
-\overline{\alpha}\ac\right)}$ is {\em displacement group} and $X_\Phi$ is the
quadrature with phase $\Phi\in[0,2\pi]$. 

The measuring apparatus is a homodyne detector with tunable phase with
respect to the local oscillator. The function
$\sqrt{2}\omega(\Phi,\sqrt{2}\lambda)$ is the probability density (with
respect to $d\lambda$ ) to obtain the value $\lambda$
measuring the quadrature $X_\Phi$, chosen randomly according to the measure
$\frac{d\Phi}{2\pi}$. Moreover, the explicit form of the estimator of
$A$, being $A$ of trace class, is
$$\sigma(A)(\Phi,\sqrt{2}\lambda) = \frac{1}{2}
\int_0^{\infty} \Tr{A e^{-it(X_\Phi -\lambda)}} t dt.$$
One could consult \cite{vasilyev} for an example of an experimental
realization of the above tomographic method.

\bg{remark}\label{oltre}
In this example one is able to obtain an estimator also for for
monomials in $\ac$ and $\cc$, \cite{R6,R7}. For example, one has that
$$\sigma(\cc\ac)(\Phi,\sqrt{2}\lambda)= 2 \lambda^2 -\frac{1}{2}.$$ 
\end{remark}
\subsection{The group $SU(2)$}

Let $SU(2)$ be the group of the unitary $2\times 2$ complex matrices with
determinant $1$. It is a unimodular connected simply-connected compact
Lie group. The corresponding Lie algebra is
$$su(2)=\set{\frac{i}{2}(x\sigma_1+y\sigma_2+z\sigma_3)\ : \ x,y,z\in\runo}$$
where $\sigma_i$ are the Pauli matrices
$$
\sigma_1=\left(\begin{array}{cc}
0 & 1 \\
1 & 0 
\end{array}\right)\room
\sigma_2=\left(\begin{array}{cc}
0 & -i \\
i & 0 
\end{array}\right)\room
\sigma_3=\left(\begin{array}{cc}
1 & 0 \\
0 & -1 
\end{array}\right).
$$
In the following we identify $su(2)$ with $\runo^3$ using the basis
$(\frac{i\sigma_k}{2})_{k=1}^3$. 
Let $V=\set{(x,y,z)\in\runo^3 \ : \ \sqrt{(x^2+y^2+z^2)}<2\pi}$, 
it is known that $V$ is an open neighborhood of $0$ such that
the exponential map restricted to $V$ is a diffeomorphism from $V$
onto the open set $\exp(V)$ and the complement of $\exp(V)$ is negligible with
respect to the Haar measure of $SU(2)$. Moreover one can check that
$$d(exp)_{(x,y,z)} = 4 \frac{\sin^2(\frac{\sqrt{x^2+y^2+z^2}}{2})}
{x^2+y^2+z^2}.$$
If we choose the Haar measure on $SU(2)$ in such a way that the
constant $D$ in Eq.\eqn{lie} is $1$ one has that
\be{norma}{\inth{1}{h} = \int_V |d(exp)_{(x,y,z)}| dxdydz = 16\pi^2,}
(usually the Haar measure on compact groups is normalized to $1$).

Given $j$ such that $2j\in\mathbb N$, let $D^j$ be the irreducible
representation of $SU(2)$ acting on $\hh=\cuno^{2j+1}$. Since the group
is compact, $D^j$ is square-integrable and the space of the
Hilbert-Schmidt operators coincides with the space of all operators
$\pippo$. 

Since the measure of $SU(2)$ is normalized according to
Eq.\eqn{norma}, it is well known that the formal degree is
$d=\frac{2j+1}{16\pi^2}$, see for example \cite{borel}.

For all $n\in S^2$,  define $J_n$ as the hermitian matrix such that
$$D^j(\exp(tn)) = e^{itJ_n}\room t\in\runo.$$
Then, the quorum defined
by $D^j$ is the set of spin operators $\set{J_n \ : \ n\in S^2}$
labelled by the space $S^2$ with the measure $\frac{dn}{4\pi}$,
being $dn$ the area element of the sphere.  It is known that the
(simple) eigenvalues of each $J_n$ are $\lambda=-j,\cdots,j$ and there
exists a unitary operator $W_n$, unique up to a phase, such that
$$Q_n = {W_n}\m J_z {W_n},$$
where $J_z=J_{(0,0,1)}.$ 

Let now $A\in\pippo$, then, according to Eq.\eqn{defi1} and
taking into account that $C_2=4\pi$, the corresponding estimator is
$$\sigma(A)(n,\lambda)=\frac{2j+1}{\pi}\int_{0}^{2\pi}e^{i\lambda
t} \Tr{A e^{-it J_n}}\sin^2(\frac{t}{2}) dt,$$
where $n\in S^2$  and $\lambda=-j,\cdots,j$. Equation\eqn{final} becomes
$$\Tr{TA} = \sum_{\lambda=-j}^{j} \int_{S^2} 
\sigma(A)(n,\lambda) |\scal{W_ne_{\lambda}}{TW_{n} e_{\lambda} }|^2\ \frac{dn}{4\pi},$$
where 
$(e_{\lambda})_{\lambda=-j}^j$ is a basis of eigenvectors of $J_z$.

This example is realized experimentally by a
Stern-Gerlach machine. The quantum system is the spin degree of freedom
of an elementary particle with spin $j$ and the number
$|\scal{W_ne_{\lambda}}{TW_{n} e_{\lambda} }|^2$ is the probability to
obtain the value $\lambda$ measuring the spin along the axis $n$,
chosen randomly according to the measure $\frac{dn}{4\pi}$.


\end{document}